\documentclass[twocolumn,preprintnumbers,
showpacs,amsmath,amssymb,aps,prd,nofootinbib]{revtex4-1}
\usepackage{graphicx}
\usepackage{color}
\usepackage[utf8]{inputenc}
\usepackage[colorlinks,hyperindex]{hyperref}  
\usepackage{accents}
\usepackage{enumitem}
\hypersetup
 {
   colorlinks,%
   citecolor=blue,%
   linkcolor=blue,%
   urlcolor=blue,%
 }
 
\def\setR{\mathbb{R}}

\def\ie {i.e.}

\newcommand{\formsp}[2]{\langle \, #1,  #2 \,\rangle}

\newcommand{\sss}[1]{\scriptscriptstyle #1} 

\newcommand{\pbundle}[4]{#1(#2,#3,#4)}

\newcommand*{\bdot}[1]{\overset{\bullet}{#1}}

\newcommand{\Miv}[1]{{\color{red}{#1}}}

\begin{document}

\title[How on earth do we get $\omega\wedge B$]{Teleparallel gravity (TEGR) as a gauge theory: Translation or Cartan connection?
}

\author{M.~Fontanini$^1$, E.~Huguet$^1$, and M.~Le~Delliou$^2$}
\affiliation{$1$ - Universit\'e Paris Diderot-Paris 7, APC-Astroparticule et Cosmologie (UMR-CNRS 7164), 
Batiment Condorcet, 10 rue Alice Domon et L\'eonie Duquet, F-75205 Paris Cedex 13, France.}
\email{michele.fontanini@gmail.com\\
huguet@apc.univ-paris7.fr}
\affiliation{$2$ - Institute of Theoretical Physics, Physics Department, Lanzhou University,
No.222, South Tianshui Road, Lanzhou, Gansu 730000, P R China
} 
\email{(delliou@lzu.edu.cn,)morgan.ledelliou.ift@gmail.com}
\date{\today}
\begin{abstract}
In this paper we question the status of TEGR, the Teleparallel Equivalent of General Relativity,
as a gauge theory of translations. We observe that TEGR (in its usual translation-gauge view) does not seem to realize the generally admitted requirements for a gauge theory for some symmetry group $G$: namely it does not present 
a mathematical structure underlying the theory which relates to a principal $G$-bundle and the choice of a connection on it (the gauge field). We point out that, while it is usually presented as absent, the gauging of the 
Lorentz symmetry is actually present in the theory, and that the choice of an Erhesmann connection 
to describe the gauge field makes the translations difficult to implement (mainly because there is in general no principal 
translation-bundle). We finally propose to use the Cartan Geometry and the Cartan connection as an alternative approach, naturally arising from the solution of the issues just mentioned, to obtain a more mathematically sound framework for  TEGR.
\end{abstract}
\maketitle
\tableofcontents
\section{Introduction}
In the present paper we are interested in the formulation of the Teleparallel Equivalent to General Relativity (TEGR) as a gauge theory.  
Let us recall that TEGR is a theory in which all the effects of gravity are encoded in the torsion tensor, the curvature being 
equal to zero: a feat achieved by choosing 
the Weitzenbock connection instead of the Levi-Civita connection of General 
Relativity (GR) \cite{Hayashi:1979qx}.  The dynamical equations for TEGR can be obtained from its action as usual (without reference to a gauge theory), thus displaying a classical equivalence with GR thanks to the fact that the Einstein-Hilbert and TEGR actions only differ by a
boundary term \cite[see for instance][]{Bamba:2013jqa}. A very important point is that TEGR is often presented as the gauge theory of the translation group  \cite{Aldrovandi:2013wha},
the main motivation for the gauge approach being
, as for many other works \cite[see][for a detailed account]{Blagojevic:2013xpa}, to describe gravity consistently with the three other fundamental 
forces of Nature which are mediated by gauge fields related to fundamental 
symmetries, namely (at our energy scale), 
U$(1)$, SU$(2)$ and SU$(3)$ for the electromagnetic, weak an strong 
interactions respectively.
By contrast with gauge theories of particle physics, in which a symmetry group 
acts in a purely internal way, the translation group, subgroup of Poincar\'e group and part of the symmetries underlying gravity, 
acts directly on spacetime and thus corresponds to an external symmetry. This 
aspect is reflected in the presence of the so-called soldering 
property\footnote{A notion 
first formulated mathematically by C. Ehresmann in 
the theory of connections \cite{Ehresmann}, a first comprehensive exposition of which can be found in 
Kobayashi \cite{Kobayashi:1957}. }, which requires 
adapting the structure of the translations gauge theory to account for it. 
Indeed, such adjustment is far from trivial, it requires 
some adaptation of the underlying mathematics and 
is also present in the larger perspective of gauge theories of gravitation. 
In the latter theories, different proposals for gauging Gravity using different symmetry groups and connections have been built without reaching
a complete consensus on the status of these proposals  \cite[see for instance][]{Sardanashvily:2016jhw,Catren:2014vza,Blagojevic:2013xpa,Tresguerres:2012nu,Wise:2006sm,Hehl:1994ue,Ivanenko:1984vf}.

The purpose of the present work is twofold: first, we will point out some difficulties in interpreting TEGR as a gauge theory of translations alone from a mathematical point of view, 
and connect these difficulties to the choice of the gauge field as an Ehresmann type connection; second, we will propose the introduction of another type of connection, known as Cartan connection, to obtain a consistent framework.

As physicists, we realize that the mathematical notions involved in the treatment of the 
topics above could be outside the common background in differential geometry. We thus  
made our goal to keep a pedagogical view throughout this work, especially when sharp distinctions between related notions are required (in particular, the distinction between the soldering and the canonical one-forms).

The paper is organized as follows. We begin in Sec. \ref{SEC-1-PreliminaryEheresmannSolderForm} with a review of the useful
mathematical structures. This section can be skipped at first reading by geometrically informed readers. In Sec. \ref{SEC-2-ConvView} we motivate our questioning about the usual formulation of  
TEGR as a gauge theory of the translation group. Then, in Sec. \ref{SEC-TranslationGauging}, we
make a comparison between the usual translation gauge theory, as exposed in \cite{Aldrovandi:2013wha}, and  a ``naive'' attempt 
to gauge the translation group following the standard mathematical point of view of connections in a principal fiber bundle, 
and conclude that the interpretation of the theory described in  \cite{Aldrovandi:2013wha} as a gauge theory of the translations is difficult to defend.  The role of connections is examined in Sec. \ref{SEC-CommentConnection} in order to motivate the use of the Cartan connection. The latter is introduced in Sec. \ref{SEC-CartanConn}, in particular through its
differences with the Ehresmann connection. As a conclusion in Sec. \ref{SEC-Conclusion} we propose to use the Cartan connection for TEGR  
and discuss its status as a gauge theory compared to other works. 
Various technicalities are described in Appendix \ref{App-A}, and for definitions not explicitly stated we refer to \cite{Fecko:2006,Nakahara:2003,KobayashiNomizu:1963}.

\section{Some preliminary notions}\label{SEC-1-PreliminaryEheresmannSolderForm}

Throughout the paper we denote by $\pbundle{P}{F}{M}{\pi}$ a fiber bundle with total space $P$, typical fiber $F$, four 
dimensional differentiable base manifold $M$ and projection $\pi$. 
Most of the time we will consider a principal $G$-bundle, that is a bundle whose fibers are identical to the structure group $G$ 
of the bundle, which in turn is a Lie group \cite{Nakahara:2003}. 
In fact, in order to be principal, a fiber bundle has to be defined along with an action of the group over the total
space \cite[p. 50]{KobayashiNomizu:1963}. It can then be shown that equivalently one can build a principal bundle from a Lie group $G$ (the fiber) and its transitions functions \cite[prop. 5.2]{KobayashiNomizu:1963}.

\begin{figure}[ht]
\begin{center}
\includegraphics[width = 1\columnwidth]{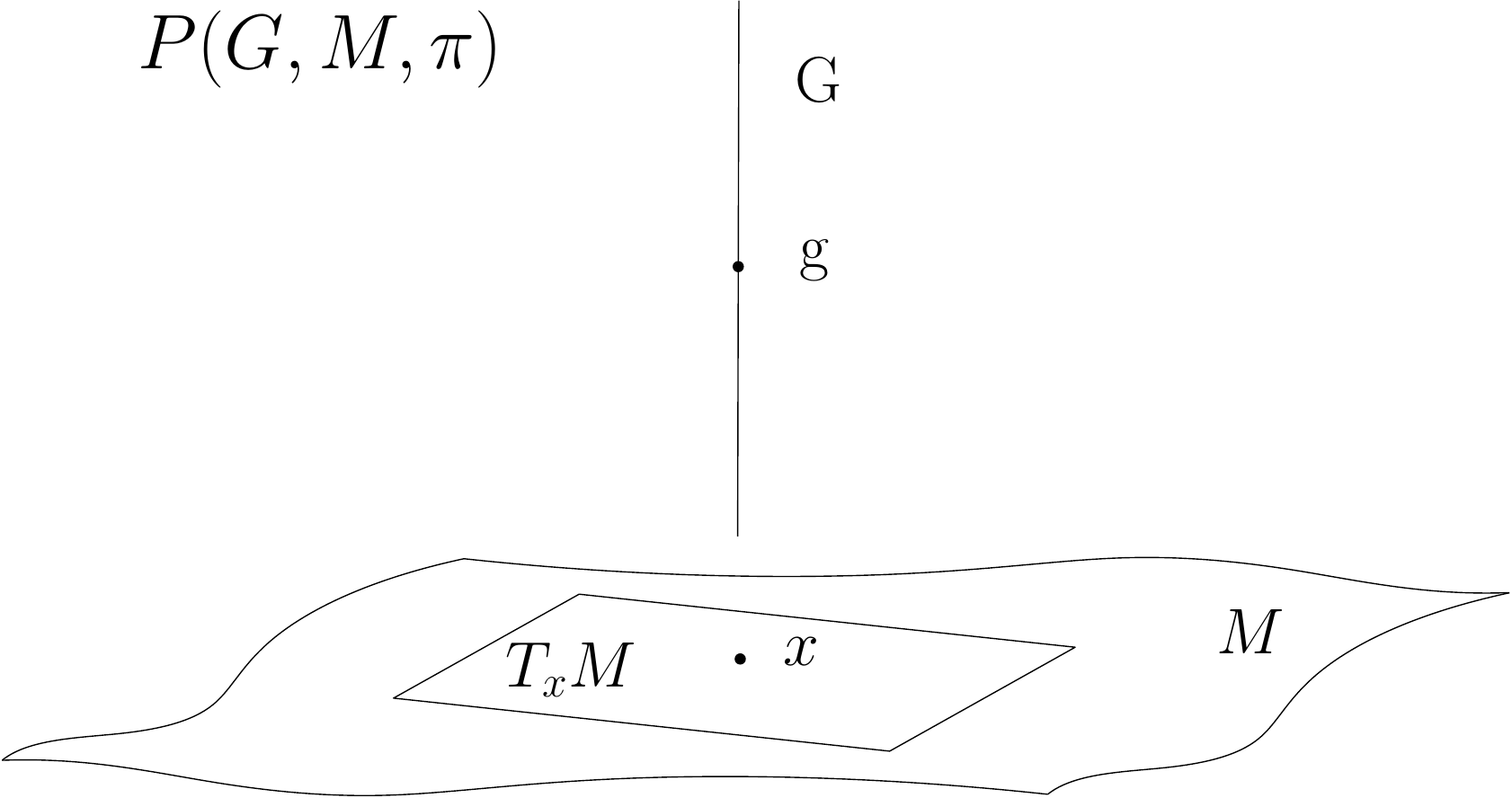}
 \caption{Generic Fiber Bundle structure}\label{FIG-FiberBundle}
 \end{center}
\end{figure}

The geometrical framework of usual gauge theories of particle physics or of  Einstein-Cartan Theory (in terms of tetrads, App.~\ref{App-DefTetrads}), of which  General Relativity  is a special case, is a principal bundle 
$\pbundle{P}{G}{M}{\pi}$ (see Fig. \ref{FIG-FiberBundle})
with a connection one-form $\omega_{\sss E}$ taking values in the Lie algebra $\mathfrak{g}$ of the group $G$. The $\mathfrak{g}$-valued one-form 
$\omega_{\sss E}$, the realization of a so-called Ehresmann connection, allows us to define the notion of parallel transport and of the curvature two-form (the latter therefore is a property of the connection), reading: 
\begin{equation}\label{EQ-DefCurvature}
 \Omega := d\omega_{\sss E} + \omega_{\sss E} \wedge \omega_{\sss E}.
\end{equation}
In gauge theories of particle physics, the group $G$ is a gauge group (U$(1)$, SU$(2)$, \ldots), and the connection and its curvature are respectively
the gauge potential and the field strength of the theory.  These are defined on 
the total space $P$ of the fiber bundle\footnote{they are often denoted by $\cal{A} \equiv$ $\omega_{\sss E}$ and $\cal{F} \equiv$ $\Omega$.},  
their corresponding quantities $A, F$ on the base manifold $M$ 
are obtained through (the pullback of) a local section $\sigma$, 
which corresponds to a choice of gauge. Explicitly~: $A = \sigma^* \omega_{\sss E}$, $F = \sigma^* \Omega$.

In the Einstein-Cartan theory the bundle considered corresponds to the orthonormal frame bundle, i.e. the bundle of orthonormal frames\footnote{Throughout the paper we assume the theory metric, the frames are always orthonormalized with respect to this unspecified metric, and so are the well known tetrads.}: each fiber above some point $x \in M$, of the base manifold 
is constituted by all orthonormal basis of the tangent space T$_xM$.  These fibers are therefore isomorphic to the Lorentz 
group\footnote{Since we are using orthonormal frames, the bundle used is a restricted frame bundle in which the fiber is the Lorentz group instead of the general linear group.}, the isomorphism being realized by choosing a specific standard frame $e$ in a neighborhood of each point $x$ and by identifying the transformed frame $e'$ with the unique element of the group $g$ realizing the transformation from $e$ to $e'$. 
The presence of such isomorphism is a necessary condition in order that the frame bundle be 
a principal fiber bundle.The connection $\omega_{\sss E}$ 
is in this context called the spin or Lorentz connection and we will denote it hereafter by $\omega_{\sss L}$.  
This Lorentz connection is related to the connection coefficients $\Gamma_{\mu\nu}^\rho$ through its pullback along some (local) section $\sigma$ by \cite[see for instance,][Sec. 15.6 and 19.2]{Fecko:2006}:
\begin{equation*}
 (\sigma^* \omega_{\sss L})_{b\mu}^a =  e_\rho^a \partial_\mu e_b^\rho + e_\rho^a \Gamma_{\mu\nu}^\rho e_b^\nu
 = e_\rho^a \nabla_\mu e_b^\rho  .
\end{equation*}
We will denote hereafter $\omega_{\sss W}$  the Lorentz connection corresponding to TEGR, i.e. the so-called Weitzenbock connection, satisfying:
$ (\sigma^* \omega_{\sss W})_{b\mu}^a = \Lambda(x)_{b}^c \partial_\mu  \Lambda_{c}^a$, where $\Lambda$ indicates a 
Local Lorentz transformation.  Note that General Relativity is obtained choosing the usual Levi-Civita connection.

Let us recall that, although the Cartan (tetrads) formalism uses a Lorentz connection in a principal fiber bundle to describe GR,  it cannot be considered a proper gauge theory of gravity. 
This is because, while local Lorentz invariance is described by the Lorentz connection, the 
so called diffeomorphism invariance of GR (the invariance under $\setR^4$ diffeomorphisms, in other words the coordinate change invariance) is not encoded in that Lorentz connection. Indeed, the description of diffeomorphism invariance 
is the main difficulty for gauge theories of gravity, which, for the most part, differ by the treatment of local Lorentz invariance and diffeomorphism invariance.

\begin{figure}
\begin{center}
\includegraphics[width = 1\columnwidth]{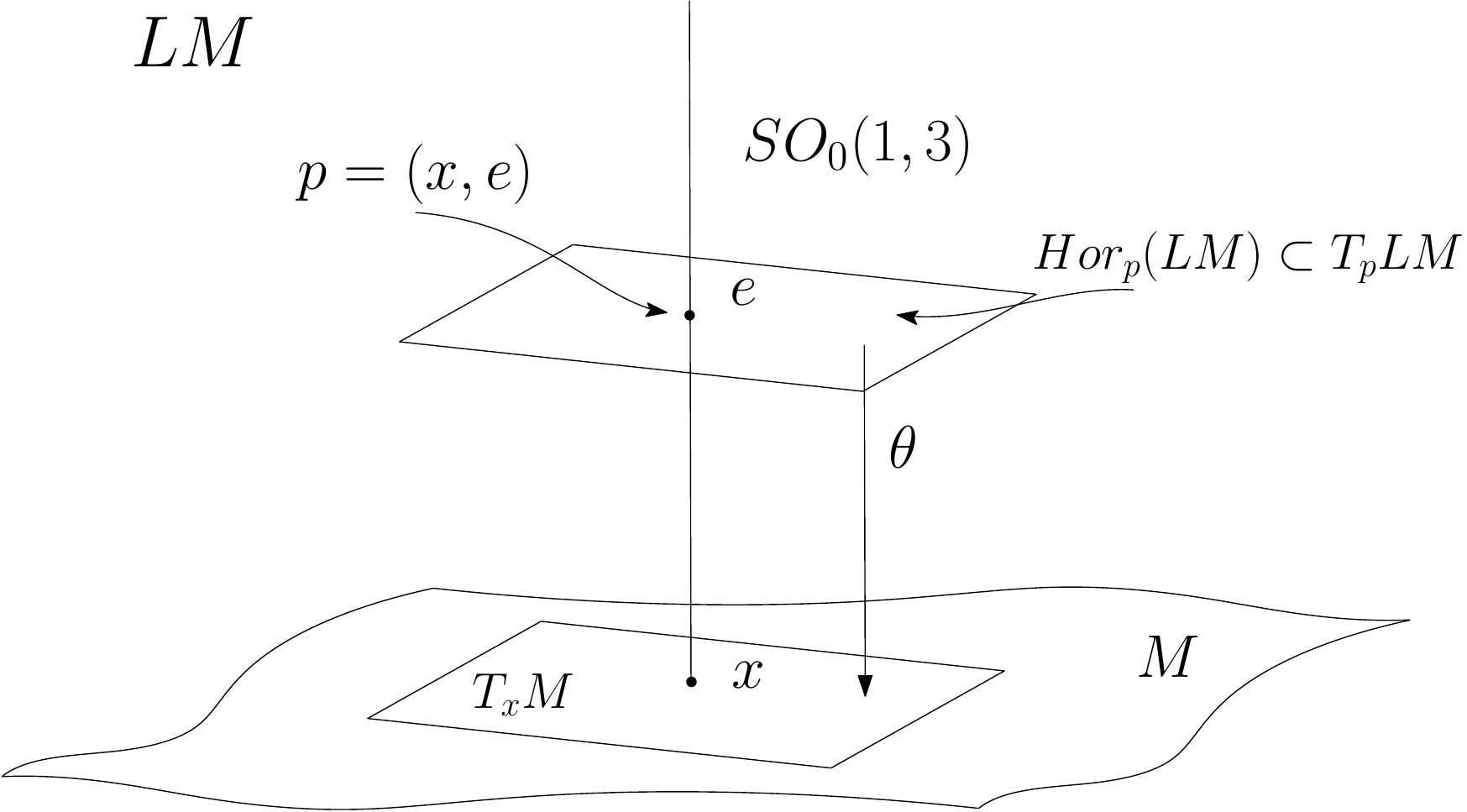}
 \caption{Frame Bundle structure and the canonical form}\label{FIG-FrameBundle}
 \end{center}
\end{figure}

It is worth noting that there is an important specificity in the choice of the frame bundle used in Tetrads formulation of General Relativity or Einstein-Cartan theory, compared to the principal bundles 
of gauge theories of particle physics, which comes from a structural difference of the theory 
and leads to the definition of torsion. A point $p$ in the frame bundle is 
basically a point $x = \pi(p)$ on the base $M$ together with a particular frame $e$ of the tangent  space $T_{\pi(p)}M$: $p=(x, e)$. As a consequence, at each point $p$ wecan obtain the components of a vector of 
$T_{\pi(p)} M$ in that specific frame $e$ at $p$. This map is realized by the canonical form $\theta$ (also named fundamental or tautological), a one-form defined on the frame bundle with 
values in $\setR^4$  relating a vector of $T_p P$ to the components of its horizontal part\footnote{Here, ``horizontal''  means  
that $\theta(v) = 0$ for $v\in\mathrm{Ver}_pP$ (the tangent space of the fiber at $p$), no connection is defined yet.} 
in the particular frame $e=\{e_a\}:\formsp{\theta^a(x,e)}{v}:=(\pi_\ast v)^a$ (see Fig. \ref{FIG-FrameBundle}).

\begin{figure}[ht]
\begin{center}
\includegraphics[width = 1\columnwidth]{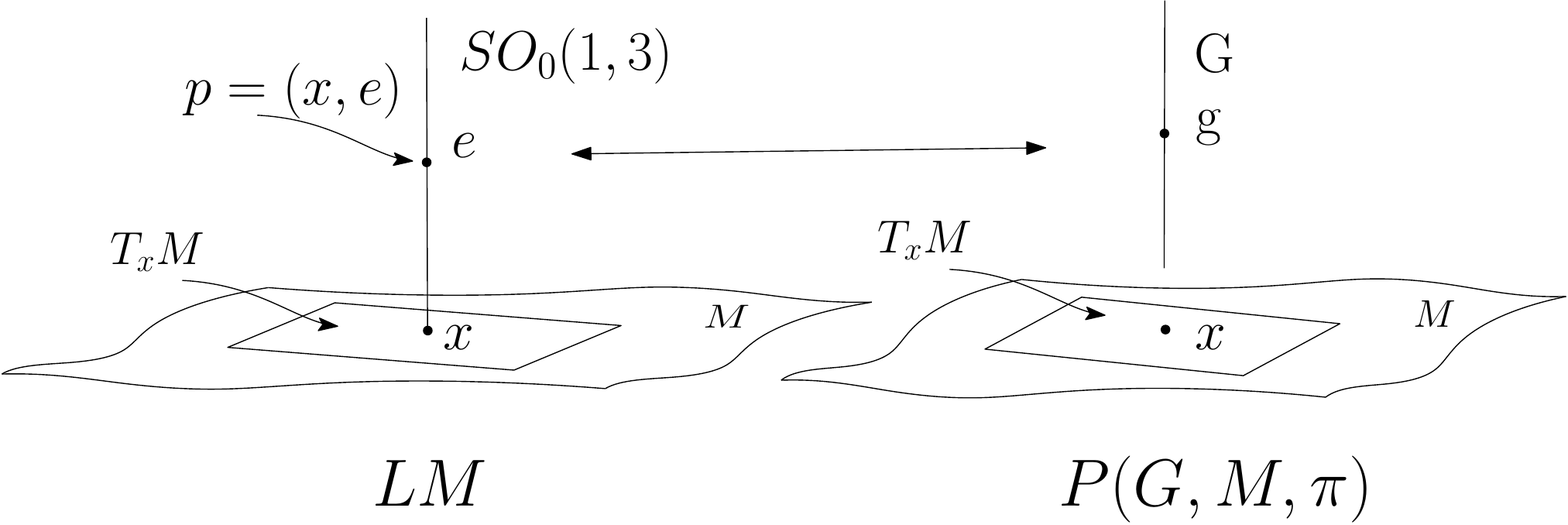}
 \caption{Frame Bundle structure (left), and a usual  G-bundle of particle physics (right) }\label{FIG-Fiber2Frame}
 \end{center}
\end{figure}

On the contrary, in the principal G-bundle of gauge theories of particle physics the frame $e$ is replaced by a ``generalized'' frame  which has nothing to do with the frames of the tangent space $T_{\pi(p)} M $  (defined in the usual way), so then there is no natural (canonical) correspondence between the two sets of frames, although we can illustrate  their similarities as 
 in Fig.\ref{FIG-Fiber2Frame}.

Therefore, when an Ehresmann connection\ $\omega_{\sss E}$  is defined on the 
frame bundle\footnote{When the considered bundle is the frame bundle the connection is said
to be linear \cite{KobayashiNomizu:1963} p. 119.}, the canonical form leads to the torsion $ \Theta $ which is defined as its
exterior covariant derivative relative to $\omega_{\sss E}$: 
\begin{equation}\label{EQ-DefTorsion}
 \Theta := d\theta + \omega_{\sss E} \wedge \theta, 
\end{equation}
$d$ being the exterior derivative 
on the frame bundle. The torsion $T$ on the base manifold is again obtained by (the pullback of) a local section $\sigma$, corresponding to a frame choice. Explicitly~: $T = \sigma^* \Theta$.  Note finally that, since choosing a  section in the 
frame bundle corresponds to choosing a frame field, one can show \cite[see for instance][Sec.21.7]{Fecko:2006}  that 
\begin{equation}\label{EQ-LinkCanoTetrad}
\sigma^*\theta^a = e^a.
\end{equation}

\begin{figure*}
\begin{center}
\includegraphics[width = 1\textwidth]{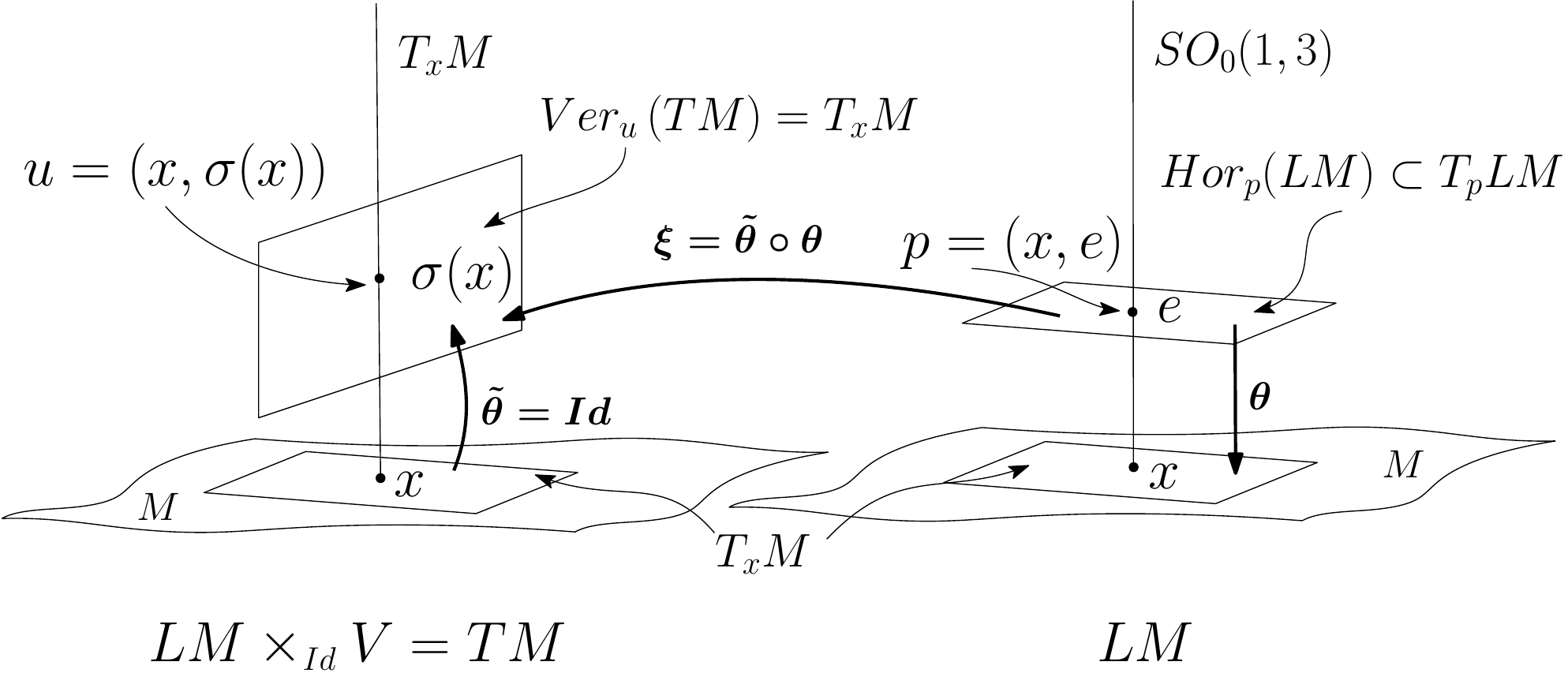}
\caption{The solder form $\widetilde{\theta}$  in the frame bundle $LM$ is the identity between the base's tangent planes 
 and the vertical spaces of the associated vector bundle, the tangent bundle, which is again the tangent plane. Note that the solder 
 form is a map between bundles ($TM\to Ver(P)=TM$ here).}\label{FIG-CanoSolderReal}
 \end{center}
\end{figure*}

The role of Eq. (\ref{EQ-LinkCanoTetrad}) is to show
 that the canonical form $\theta$ realizes the so-called  ``soldering'' between the base manifold and the fibers. It is important though to distinguish the canonical form $\theta$ from the soldering form $\tilde\theta$ which is a different mathematical object (see appendix \ref{App-SolderForm}).  
Basically, the soldering for a principal bundle
identifies each tangent space $T_xM$ of the base manifold at $x$ with a corresponding space $T_{\sigma(x)}V$, 
tangent to a fiber of an associated vector 
bundle --~that is, in short, a bundle in which the principal bundle fiber $G$ is replaced by a representation 
of $G$ on a vector space $V$ (see appendix \ref{App-AssocBundle})~-- along a global section $\sigma$, as discussed in appendix \ref{App-SolderForm}.
In the case of the frame bundle formulation of GR or Einstein-Cartan theory,  the tangent bundle $TM$  is itself an associated vector bundle, and as such, the tangent space to a fiber of the 
associated vector bundle and the tangent space of the base manifold of the principal bundle's base are the same, and the 
soldering form is from this point of view the identity\footnote{Hence the name ``tautological form'' for the canonical form $\theta$.}%
as we illustrate in Fig.\ref{FIG-CanoSolderReal}.
It is worth noting there that the solder and canonical forms are distinct and that the principal bundle soldering, effected through the isomorphism $\xi$, uses the composition of the two, by which we can henceforth understand that, since here the soldering form is the identity, the canonical form realizes the soldering.\\

\section{Some questions about the usual translation gauge formulation}\label{SEC-2-ConvView}
It is often claimed that TEGR can be formulated as a gauge theory for the translation group. Nevertheless,  as noted by Aldrovandi and 
Pereira \cite[p.41]{Aldrovandi:2013wha}, since ``\emph{The gauge bundle will then present the soldering property \ldots Teleparallelism will be necessarily a non-standard gauge theory}''. In the present section we do not enter in the 
mathematical details of the gauge formulation of TEGR but, closely following \cite{Aldrovandi:2013wha}, in which this gauge approach is comprehensively described, we point out some of these ``non-standard'' aspects 
which, in our opinion, raise questions about the interpretation of TEGR as a gauge theory of the translation group. 
Although we follow the presentation of \cite{Aldrovandi:2013wha}, we prefer to use the language 
of differential geometry, which from our point of view allows us to express our arguments in a more concise way.

Similarly to other gauge theories, in TEGR the effects of the local symmetry (gauge) group, the translation group, are implemented through the gauge potential denoted $B$ from here on. 
The field strength of $B$, hereafter denoted by $\bdot{T}$, is given in  Eq. (4.52) of \cite{Aldrovandi:2013wha}. Using  the definition of
exterior derivative and wedge product it reads 
\begin{equation}\label{EQ-TfromB}
\bdot{T} = dB + \omega_{\sss L} \wedge B,
\end{equation}
$\omega_{\sss L}$ being the Lorentz connection for the theory.  That expression is recognized as 
being equal to the torsion of the connection $\omega_{\sss L}$.

In Ref.\cite{Aldrovandi:2013wha} we find another expression for $\bdot{T}$ that connects it to tetrads (Eq. 4.62)
\begin{equation}\label{EQ-Tfromh}
\bdot{T} = dh + \omega_{\sss L} \wedge h,
\end{equation}
where $h$ is a, possibly nonholonomic, tetrad. Moreover, the two expressions (\ref{EQ-TfromB}) and (\ref{EQ-Tfromh}) are linked by a relation between $B$ and $h$ \cite[][Eq.4.47]{Aldrovandi:2013wha}
\begin{equation}\label{EQ-LinkTetradsBandh}
h = e + B,
\end{equation}
where $e$ is a tetrad satisfying the so--called Maurer--Cartan formula:~$de + \omega_{\sss L}\wedge e = 0$. 

Interpreting the field strength $\bdot{T}$ as curvature of the gauge field $B$ is definitely non--standard for 
gauge theories. In fact, comparing (\ref{EQ-TfromB}) and (\ref{EQ-DefTorsion}) we note that the field strength $\bdot{T}$ matches the general definition of torsion of the Lorentz connection $\omega_{\sss L}$, provided $B$ plays the role 
of a tetrad, and the geometrical structure describing the theory includes the frame bundle on which $\omega_{\sss L}$ is defined. Let us stress the fact that the point of this ``non-standardness'' is not to make the curvature of $B$ equal to the torsion of  $\omega_{\sss L}$, 
but rather, in order to ensure this equality, to make $B$, which as a gauge field of the translation group is a connection 
one form (up to a pull back) in the translation-bundle (whatever it may be),
also be a tetrad in the frame bundle.

What we just described exemplifies well the kind of difficulties we wish to point out.
We stress out that \cite{Aldrovandi:2013wha} uses mostly the tensorial formalism, and in particular 
the ``dual'' definition of the field strength, given 
by the commutator of the covariant derivatives related to the gauge field:  $\bdot{T}_{\mu\nu} = \left[h_\mu, h_\nu\right]$, 
the tetrad $h$ being there, as a differential operator, the covariant derivative. As a consequence, 
the non--standard use of the objects seen above appears more natural.

Finally, another question raised by the translation gauge approach is that: whilst here the Lorentz group is described as left ungauged, the expression of the covariant derivative  $h$ is the same as that appearing in theories obtained from a very different geometrical structure in which the whole Poincar\'e group is taken as the gauge group
, thus including the Lorentz local symmetry. For instance in Eq.  (72) of Tresguerres's work
\cite{Tresguerres:2012nu}.

We conclude from the above remarks that the link between the physical quantities and their mathematical counterpart
is at least puzzling. For instance, one may ask if, in the ``non-standard'' translation gauge approach, the field strength must always be described by the curvature of some connection? On the other hand, what is the 
meaning of gauging two different groups, with two different (connections) gauge fields,  leading to the same covariant derivative (the tetrad $h$)?  

\section{A conventional gauging of the Translation group}\label{SEC-TranslationGauging}

We now compare the standard objects (connection, curvature, \ldots) and their physical 
counterparts (gauge field, field strength, \ldots) that appear in the mathematical framework of principal fiber bundle\Miv{s} used to describe gauge theories
\cite[see for instance Ch. 21 of][]{Fecko:2006}, with the ``non-standard'' translation gauge approach described in \cite{Aldrovandi:2013wha}, and 
discussed above. To make the comparison explicit, we build a ``naive'' translation gauge theory following the general framework applied to the  translation group.
As a word of caution, we stress out that this 
theory does not pretend to describe a viable gauge theory of translations, in particular because, as we will see, the translations bundle cannot in general be principal. Indeed, this is merely an exercise 
allowing us to pinpoint where the standard treatment and the works in Ref.~\cite{Aldrovandi:2013wha} diverge, to explain why, and how, in our point of view, the latter theory status as a gauge theory of translation is difficult to defend.

Table \ref{TAB-TranslGauge} displays the general mathematical ingredients of the models in the cases of a general fiber bundle theory, its specifications to the hypothetic principal translation group bundle (whatever it may be), 
and  (if any) the equivalent objects in the ``non-standard'' translation 
gauge approach of \cite{Aldrovandi:2013wha}. 
\begin{table}
\begin{tabular}{|l|l|l|}
\hline
General&$G=T_4$&TEGR from Ref.~\cite{Aldrovandi:2013wha}\\
\hline
\hline
$\pbundle{P}{G}{M}{\pi}$ & $\pbundle{P_{\sss T}}{T_4}{M}{\pi}$ & $TM$ \\
$(x,f)$& $(\{x^\mu\},\{v_g^\alpha\})$ & $(\{x^\mu\},\{x^a\})$ \\
 $\omega=\omega^{\sss I}E_{\sss I}$  & $\omega_{\sss T}= \omega_{\sss T}^a P_a$ & ~\\
 $\Omega=d\omega $ & $\Omega_{\sss T}=d\omega_{\sss T} $&~\\
$+ \omega\wedge\omega$& $+ \omega_{\sss T}\wedge\omega_{\sss T}$& ~\\
 $\sigma$&$v_g(x)$ & $\{x^a(x)\}$\\
 $A=\sigma^*\omega$ & $A_{\sss T} = \omega_{\sss T}(v_g)$ & $B$\\ 
 $F=\sigma^*\Omega$ & $F_{\sss T} = d A_{\sss T} $ & $\bdot{T} = dB + \omega_{\sss L} \wedge B$\\
\hline
\end{tabular}
\caption{Comparison of fiber bundle/gauge theory objects. The first column contains the main elements 
in a general fiber bundle approach. In the second column the gauge group is specified to translations ($T_4$). The last column 
shows the equivalent objects  for the ``non-standard'' approach of  \cite{Aldrovandi:2013wha}.}\label{TAB-TranslGauge}
\end{table}
In  the first line we have the specifications of the corresponding fiber bundle. The fiber bundle of \cite{Aldrovandi:2013wha} 
is identified, following the authors, as the tangent bundle. 
The local trivializations, in the general case $(x,f)$, are presented in the second line. 
The translation group case can be paralleled to the frame bundle identification between a frame and 
a Lorentz group element, by identifying a translation $g$ with the corresponding vector $v_g$ in $\setR^4$.
For the authors in \cite{Aldrovandi:2013wha}, this corresponds to tangent plane coordinates $x^a$.
The connections $\omega$ and their associated curvature $\Omega$ are presented for the general case, with Lie algebra 
basis $\{E_{\sss I}\}$ and, for the $T_4$ case, with the abelian translation algebra basis $\{P_a\}$.
The section $\sigma$ is a vector field $v_g(x)$ for $T_4$, while as in \cite{Aldrovandi:2013wha} it is 
given (in components) by $x^a(x^\mu)$. The connection's curvature on the spacetime (base) is noted $F$, given 
by the curvature $\Omega$'s pullback along $\sigma$, and reduces to $dA_{\sss T}$ for our ``naive''  translation theory since the translation group is abelian, while the ``non-conventional'' approach uses Eq. (\ref{EQ-TfromB}).
We thus see that, apart from
a change in notation and the use of the definition of curvature in the total space not
presented in \cite{Aldrovandi:2013wha}, the differences between the ``naive'' and the usual theories are on the nature 
of the bundle of translations itself  \cite[identified with $TM$ in][]{Aldrovandi:2013wha}, and the expression of the field strength  ($\bdot{T}$ vs $F_{\sss T}$). These two points deserve a deeper comment based on some important 
mathematical details \cite[see][]{Fecko:2006, Nakahara:2003, KobayashiNomizu:1963}.

The usual mathematical description of a gauge theory of translations would adopt a principal fiber bundle. 
That entails the bundle should employ the translations group in two roles:
as structure group and as (being isomorphic to) fibers.  Now, the tangent bundle without additional specification is a vector bundle whose structure group is, 
for a $n$-dimensional real base manifold, GL$(n, \setR)$, the general linear group\footnote{We consider the general $n$-dimensional case in the present paragraph.}. 
However, the linear group GL$(n,\setR)$ does not even contain the group $T_n$ of 
$n$-dimensional translations. In addition, as a vector bundle, each  fiber of $TM$ is the vector space $\setR^n$, 
which is also, as a manifold, the translation group $T_n$. However, in order to identify the translation
group with the fiber in the principal translations bundle one would first have to exhibit a right action of that group 
on the total space, an action which is not defined in the tangent bundle and in any case would not use the structure group GL$(n, \setR)$. 

One could ask if there is a possibility to associate in a natural way a principal bundle with the tangent bundle. The answer 
is yes but it turns out that this bundle is precisely the frame bundle, and no translations are present there. 
Indeed, the problem of defining an action on the whole manifold forbids the tangent bundle, viewed as a translation-bundle - that is, whose fibers are the translation group - to be principal.  This is because, if one considers 
an arbitrary vector of $\setR^n$, viewed as the (manifold of the) translations
group, one must define its action on the total space of $TM$, viewed as the bundle of translations.  To this end, one has to 
specify how to identify $\setR^n$ to each tangent space $T_x$ of $TM$, that is to specify a frame for each $T_x$. Put in another way we need a field of frame on the whole base manifold, which is precisely a global section of the frame bundle. Now, there is a theorem which states that a principal bundle (as the frame bundle) admitting
a global section is trivial \cite[see for instance,][Sec. 20.1]{Fecko:2006}. Thus, the bundle of translations is principal if and only if the associated frame bundle is trivial. Note that, this argument says at once that the tangent bundle is not in general a principal $T_n$-bundle and that the hypothetic principal $T_4$-bundle is not defined in general.

Let us  now consider the  other main ``non--standardness'',  
namely the expression for the curvature that in \cite{Aldrovandi:2013wha} contains an additional 
term $\omega_{\sss L} \wedge B$ with respect to the ``naive'' version.
We first note that the latter, built by gauging translations alone, cannot account for the local Lorentz invariance and  thus 
fails to properly describe gravity. Therefore a  heuristic way to implement local Lorentz invariance would be to use a minimal coupling procedure, leading to the replacement of the exterior derivative $d$ by its Lorentz covariant 
counterpart ($D := d + \omega_{\sss L} \wedge$) in the expression for the curvature, eventually giving
precisely the  ``non-standard'' approach expression (\ref{EQ-TfromB}). Indeed, this would be 
coherent with the view adopted in \cite{Aldrovandi:2013wha} in which one considers $B$ as the gauge field for the translations while the local Lorentz invariance is related to the non-holonomic frames. Moreover, this approach points towards the interpretation of the ``dual''  role of the field $B$ as a tetrad, which in this view is somehow forced by the expression of the torsion (\ref{EQ-DefTorsion}), related to the canonical form through (\ref{EQ-LinkCanoTetrad}). The only concern with this heuristic view is that the introduction of a covariant derivative corresponds, on mathematical grounds, to gauging the 
Lorentz symmetry.

The discussion above
allows us to clarify the doubts raised at the end of the previous section about the fact that the 
theory of translations presented by Aldrovandi and Pereira in \cite{Aldrovandi:2013wha} and the gauge 
theory of Poincar\'e  symmetries proposed by Tresguerres in \cite{Tresguerres:2012nu} lead to the same expression for the covariant derivative. Indeed, what is happening is that in  \cite{Aldrovandi:2013wha} the Lorentz symmetry is implicitly gauged. The fact that the Lorentz connection is introduced in order
to take into account the most general orthonormal frames (including non-holonomic ones), does not allow us to  
ignore its mathematical nature, namely, a connection in the orthonormal frames bundle. From this point of view, the heuristic introduction of the Lorentz covariant 
derivative (the replacement of $d$ by $d + \omega_{\sss L}\wedge$) in the previous paragraph to account for local  Lorentz invariance in a gauge theory of translations 
is somehow reminiscent of the use of the composite connection in Tresguerres works  \cite{Tresguerres:2012nu}.

To summarize, although the ``non-standard'' gauge theory 
described in \cite{Aldrovandi:2013wha} 
reproduces TEGR, in our opinion it is difficult to consider it a gauge theory of the translation group. 
Our view is mainly motivated by the fact that the bundle of translations,  whatever it may be,  
cannot be identified, as a principal $T_4$-bundle, with the tangent bundle. In addition, the claim that the 
Lorentz symmetry is left ungauged seems, on mathematical grounds, at odds with the definition of a covariant derivative that includes a term for a Lorentz connection. In our view, the usual ``non-standard'' gauge theory of translations presented as TEGR is the Einstein-Cartan (tetrad) formulation of gravity, which takes place in the principal orthonormal frame bundle, with the Lorentz connection chosen to be the Weitzenbock connection; the translational part, which does not arise from a gauge in the Einstein-Cartan formulation,  is given there by the canonical form $\theta$, viewed as the translational 
(part of the) connection whose pullback on the base manifold through Eq. (\ref{EQ-LinkCanoTetrad}) is the gauge field $B$.

\section{Some comments about the connections}\label{SEC-CommentConnection}
The commonly accepted mathematical framework for gauge theories is that of principal bundles in which local gauge symmetries correspond
to connections. 
In the case of gravitation, the local symmetries are
\begin{enumerate} 
\item the local Lorentz invariance. and
\item the invariance under the local ($\setR^4$) diffeomorphism, which corresponds to local translational invariance.
\end{enumerate} 
As already mentioned in 
Sec.~\ref{SEC-1-PreliminaryEheresmannSolderForm}, to account for these diffeomorphisms from a gauge perspective is a central difficulty of
gauge theory of gravity, and translates into an equally difficult
choice of the connection. The present paper being devoted to the gauge version of TEGR, we do not aim to discuss general gauge theories of gravity. We nevertheless comment about the connections with the aim to motivate our proposal to  
use the Cartan connection in next section.

On one hand, as shown in Sec. \ref{SEC-TranslationGauging}, \cite{Aldrovandi:2013wha}'s exposition of a ``non-standard'' gauging of translations does not fit well in the principal fiber bundle mathematical framework that should describe a translations gauge theory, as there 
\begin{enumerate}
\item translations are not properly taken into account as a connection in a principal bundle,
\item the Lorentz symmetry is implicitly gauged.
\end{enumerate}
On the other hand, a more straightforward approach in which only a gauge field for the translations would be considered fails because 
\begin{enumerate}
\item the local Lorentz invariance is not satisfied by a translational field alone
\item a fiber bundle for the translations moreover fails to be principal except if it is trivial.
\end{enumerate}

It thus seems that the two invariances, local Lorentz and local translations ($\setR^4$ diffeomorphisms) should be considered together in a connection allowing  TEGR to be accounted for as a gauge theory. 
Since the force field strength for TEGR, i.e. the bundle's curvature (up to a pullback), is the torsion, such a connection would have to yield torsion as its curvature. Moreover, being that TEGR lives in manifolds with no curvature, a corresponding Weitzenbock connection should be the choice for its Lorentz connection.
One can propose a simple ansatz for 
a connection, say $\omega$, satisfying the properties just discussed. It can be written as (omitting all indices to 
keep matter simple) 
\begin{equation*}
\omega = \omega_{\sss L} + \theta_{\sss T}, 
\end{equation*}
where $\omega_{\sss L}$ is a Lorentz connection and $\theta_{\sss T}$ embodies
the translational part. 
The corresponding curvature then reads
\begin{equation}\label{EQ-Curvature_Om+Thet_Generic}
 \Omega :=  d\omega + \omega\wedge\omega
= \Omega_{\omega_{\sss L}} + \Theta_{\omega_{\sss L}} + \theta_{\sss T} \wedge\omega_{\sss L}, 
\end{equation}
where $\Omega_{\omega_{\sss L}} := d\omega_{\sss L} + \omega_{\sss L}\wedge\omega_{\sss L}$ and 
$\Theta_{\omega_{\sss L}} :=  d\theta_{\sss T}  + \omega_{\sss L}\wedge\theta_{\sss T}$, are respectively the curvature and 
the torsion of the Lorentz connection. Note that the term $\theta_{\sss T}\wedge\theta_{\sss T}$ is zero since the group of translations
is abelian.

The curvature of the connection $\omega$, as shown in Eq. (\ref{EQ-Curvature_Om+Thet_Generic}), reduces to the torsion 
of the Lorentz connection if 
\begin{enumerate}[leftmargin=*]
\item $\omega_{\sss L} = \omega_{\sss W}$, the Weitzenbock connection, 
\item the last term $\theta_{\sss T} \wedge\omega_{\sss L}$ vanishes.
\end{enumerate}   Note that the same is true for the curvature on the base manifold by pulling back (\ref{EQ-Curvature_Om+Thet_Generic}) along some local section.

Now, since an Ehresmann connection on a principal $G$-bundle, takes it values in the Lie algebra of the whole group $G$, it is difficult to see how to make the cross term $\theta_{\sss T} \wedge\omega_{\sss L}$ vanish. For instance, the gauge theory of the Poincar\'e group using a composite connection proposed in \cite{Tresguerres:2012nu} leads to a  connection on the base manifold which is the sum of a Lorentz and a translational part \cite[see Eq. (70) of][]{Tresguerres:2012nu},  
but the cross term forbids  identification of that connection's curvature with the torsion of the Lorentz connection.

A (perhaps?) more natural way to implement 
translations would point to the use of an affine connection 
(see appendix \ref{App-AffineConnection}). However the affine connection using 
Lorentz and translations is not the simple sum of the Lorentz and translations connections and neither is its curvature yielding the sum 
of Lorentz curvature and torsion directly \cite[Sec.3.3 p125]{KobayashiNomizu:1963}.

Lastly, 
the Cartan connection, which is not of Ehresmann type, reduces exactly to take the form of the connection $\omega$.
In the following section therefore, we will describe the main features of this connection and then use it to obtain TEGR.

\section{Approaching TEGR with the Cartan connection}\label{SEC-CartanConn}

The Cartan connection appears in the context of Cartan geometry which can be seen as a generalization of  
Riemannian geometry in which the tangent space is replaced by a tangent homogeneous (\ie,~maximally symmetric) space. This geometry and the properties of the Cartan connection, in relation with
gravity theories, are summarized in a comprehensive way  by Wise \cite{Wise:2006sm} and Catren \cite{Catren:2014vza}. A detailed mathematical reference is given by Sharpe \cite{Sharpe:1997}, see also 
\cite{Francois-Lazzarini-Masson:2014} for a summary and a comparison with other mathematical approaches.

As in the case of usual gauge theories of particle physics or of the Einstein-Cartan theory, 
the geometrical framework of Cartan geometry is a principal bundle $\pbundle{P}{H}{M}{\pi}$ with a 
connection $\omega_{\sss C}$.  However,  there are three important  differences with respect to the 
usual case:
\begin{enumerate}
 \item The fiber  is here a (topologically closed)  subgroup $H$ of a larger Lie group $G$, \label{enu:fiberH}
 \item The connection is a Cartan connection $\omega_{\sss C}$ which takes its values in the 
 algebra $\mathfrak{g}\supset\mathfrak{h}$ of $G$.\label{enu:cartanConn}
 \item The connection $\omega_{\sss C}$ is, at each point $p$ of $P$, a linear isomorphism between the tangent space $T_pP$ and the Lie algebra $\mathfrak{g}$. This property requires that $G$ has the same dimension as the
tangent space $T_p P$.\label{enu:gConn}
\end{enumerate}
These properties  are specific to the Cartan geometry, in particular 
(\ref{enu:cartanConn})-(\ref{enu:gConn}) distinguish the Cartan connection from Ehresmann's, 
which by contrast takes its values in the Lie algebra of the group $H$ (the fiber) and does not satisfy a property like $(3)$. As a consequence
of the above properties, the tangent space of the base manifold $M$ can be locally 
identified with the tangent space $\mathfrak{g}/\mathfrak{h}$ of the homogeneous 
space\footnote{
Note that both $G/H$, with $H$ a closed subgroup of $G$, being   
an homogeneous space, and the fact that $\mathfrak{g}/\mathfrak{h}$ 
can be identified with its tangent space are known results of differential geometry of Lie groups (see for instance \cite{Fecko:2006} p. 294  for the former statement, and \cite{Sharpe:1997} p. 163, for the latter).
}
$G/H$. Indeed, the third condition precisely states
that the principal bundle $\pbundle{P}{H}{M}{\pi}$ is soldered to the base $M$.

For a $(3+1)$-dimensional manifold there are only three possible homogeneous spaces: the (Anti)-de~Sitter spaces and the Minkowski space. Each one being also a group of symmetry:  SO$_0(2,3)$, SO$_0(1,4)$ for Anti-de~Sitter 
and de~Sitter spaces respectively and the Poincar\'e group for the Minkowski space.  The corresponding Cartan geometries have the property of being reductive; the definition of this property
can be found in  \cite[p197]{Sharpe:1997}, along with its differences with the notion of reductive algebra. From our perspective it is sufficient to say that, for a reductive Cartan geometry, the Cartan connection takes the form
\begin{equation}
 \omega_{\sss C} = \omega + \theta,
\end{equation}
where $\omega$ is an Ehresmann connection  $\mathfrak{h}$-valued one-form, on the principal fiber bundle $\pbundle{P}{H}{M}{\pi}$,  and $\theta$ a $\mathfrak{g}/\mathfrak{h}$-valued one-form on $P$.  Moreover, the definition of the reductive geometry, which relies on the existence of an isomorphism between adjoint representations of the structure group $H$ (an Ad($H)$-invariant decomposition of $\mathfrak{g}$, namely: $\mathfrak{g} \simeq \mathfrak{h} \oplus \mathfrak{g}/\mathfrak{h}$),  ensures that the two parts ($\omega$ and $\theta$) of the Cartan connection $\omega_{\sss C}$ remain separated under a gauge transformation (a change of local section in $\pbundle{P}{H}{M}{\pi}$). 

For what may concern us, the main implication of the (Ad($H)$-invariant) decomposition of the 
Lie algebra  $\mathfrak{g}$ is that it allows us to split any $\mathfrak{g}$-valued form defined on $P$. 
In particular the curvature two-form $\Omega_{\sss C}$ of the Cartan connection $\omega_{\sss C}$ which reads
\begin{equation}\label{EQ-CartanCurvatureDecomposition}
 \Omega_{\sss C} : = d\omega_{\sss C} + \omega_{\sss C}\wedge \omega_{\sss C} = \Omega_\omega + \Theta_\omega,
\end{equation}
 were $ \Omega_\omega$ and  $\Theta_{\omega}$ stand respectively for the curvature and  the torsion of the  Lorentz (Ehresmann) connection $\omega$. An explicit calculation, using the fundamental representation of $\mathfrak{g}$, is given by Wise \cite{Wise:2006sm} for the three (tangent) homogeneous spaces: Anti-de~Sitter, de~Sitter and Minkowski. For the Minkowski case one obtains, by choosing the Weitzenbock connection $\omega = \omega_{\sss W}$, the expected result~: 
\begin{equation*}
 \Omega_{\sss C} = \Theta_{\omega_{\sss W}}.
\end{equation*}

Introducing the reductive Cartan geometry thus solves the problem stated in the previous section: 
it accounts properly for both Lorentz and translational symmetries through a connection whose curvature is the torsion. 
Moreover, the specific case of reductive Cartan geometries allows us to retrieve the framework of the orthonormal frame bundle: as shown in \cite{Sharpe:1997}, for a reductive geometry the first part $\omega$ of the Cartan connection
is precisely an Erhesmann connection and the second part $\theta$ realizes the soldering (see appendix \ref{App-SolderForm}); in addition, the bundle $\pbundle{P}{H}{M}{\pi}$ is necessarily a reduction from the $GL(\setR)$ frame bundle on $M$ to the subgroup $H$ leading thus to the 
orhonormal frame bundle for $H = \mathrm{SO}_0(1,3)$.

\section{Conclusion}\label{SEC-Conclusion}

The results of the previous sections lead us to propose to view TEGR in the context of reductive Cartan geometries. The main reason is that, in our opinion, such geometries
provide a more consistent framework than the usual translation-gauge theory. 
Precisely, we argued in Sec. \ref{SEC-2-ConvView} and \ref{SEC-TranslationGauging} that 
\begin{enumerate}
\item the conventional approach makes use of a translational gauge field which does not appear, on mathematical grounds,  as a connection in a principal "translation-bundle"
\item moreover such a bundle could only be defined for spacetimes with trivial frame bundles, and finally
\item local Lorentz symmetry, taken into account in order to implement covariance properties, appears as a gauge field since it is a connection in the principal bundle of orthonormal frames.
\end{enumerate}   
Clearly, the Cartan connection corresponding to the Poincar\'e symmetry does not share 
these drawbacks and, in the case of the Weitzenbock connection of TEGR, has torsion for curvature.
In addition, the frame bundle soldering to the base manifold, recognized as a source of difficulties in the gauge theoretical context, is a "built-in" property for Cartan geometries.

On the other hand, the question about promoting TEGR to a legitimate gauge theory by building it with a Cartan connection is still open
. Indeed, the structure of the reductive Cartan geometries differs from that of 
the usual gauge theories mainly because the connection does not only relate to the symmetry group (here the Lorentz group) of the principal bundle  on which it is defined (here the orthonormal bundle of frames): the  
connection, in fact, takes its value in a larger lie algebra (here the Poincar\'e algebra). This  peculiarity could be considered as the "non-standardness" inherited from the soldering property as mentioned in \cite{Aldrovandi:2013wha}. More importantly, although this Cartan version of TEGR 
cannot be considered as a Poincar\'e Gauge Theory (PGT)\footnote{It could be part of a new class of "Cartan-Poincar\'e- Gauge Theory" (CPGT) and termed "Weitzenbock-CPGT".},  the 
whole Poincar\'e symmetry is required. This point should
be compared to the work of \cite{Itin:2016nxk}, which adopts a very different approach, but still 
reaches the conclusion that TEGR cannot be obtained without the whole Poincar\'e symmetry.

\section*{Acknowledgements}

The authors wish to thank G. Catren and M. Lachi\`eze-Rey for very useful discussions and references about differential geometry,  
and F. Helin for his help on some mathematical notions and discussions. MLeD acknowledges the financial support by Lanzhou University starting fund.

\appendix
\section{Definitions of, and comments on, some mathematical structures}\label{App-A}
\subsection{Tetrads}\label{App-DefTetrads}
We will take the definition of Nakahara's Book \cite{Nakahara:2003} (Sec. 7.8.1) that is~: the tetrads (vierbein)
are the coefficients of the (field of) basis vectors $\{e_a\}$ non-coordinates, orthonormalized and preserving orientation.   We have
$g_{\mu\nu}e_a^\mu e_b^\nu = \eta_{ab}$, $g_{\mu\nu}= e_\mu^a e_\nu^b \eta_{ab}$. 

\subsection{Comment on Ehresmann connection}\label{App-CommentOnEhresmannConn}
An Ehresmann connection on a principal G-bundle basically  provides a G-invariant splitting of the tangent space  $T_pP$ into a vertical Ver$_pP$ and 
a horizontal Hor$_pP$ part by defining (uniquely) the horizontal vectors as those which belong to its kernel. 
Although always possible, the tangent space  $T_pP$ splitting into a vertical and a horizontal part is not unique. While the vertical part is always 
uniquely defined -- the vertical vectors belong to the kernel of $\pi_\ast$, the horizontal part can be, in general, 
any complementary space of  Ver$_pP$ in $T_pP$. In this sense, both verticality and horizontality are always 
defined. In addition there is a linear isomorphism between  Ver$_pP$ and the 
Lie algebra $\mathfrak{g}$ (see for instance \cite{Fecko:2006} p. 560) and between Hor$_pP$ and $T_{\pi(p)} M$ 
through $\pi_\ast$ (see for instance \cite{Isham:1999} p. 255).  The Ehresmann connection specifies a unique horizontality which in turn is 
used to define the parallel transport of various
tensorial objects.  

Since a vector of $T_pP$ can be split in a unique way into a horizontal and a vertical part, there is a map which projects
the vector along Hor$_pP$ to Ver$_pP\simeq\mathfrak{g}$, 
that is a $\mathfrak{g}$-valued one-form. Then,  
the definition of an Erhesmann connection on a principal $G$-bundle is often made through that connection one-form $\omega_{\sss E}$ whose kernel specifies the horizontal vectors of $TP$. This definition of horizontality has to be consistent with the group action. A formal definition, together  with
a comparison with the Cartan connection, is detailed in \cite{Wise:2006sm}.

\subsection{Solder form}\label{App-SolderForm}

\begin{figure}[th!]
\begin{center}
\includegraphics[width = 1\columnwidth]{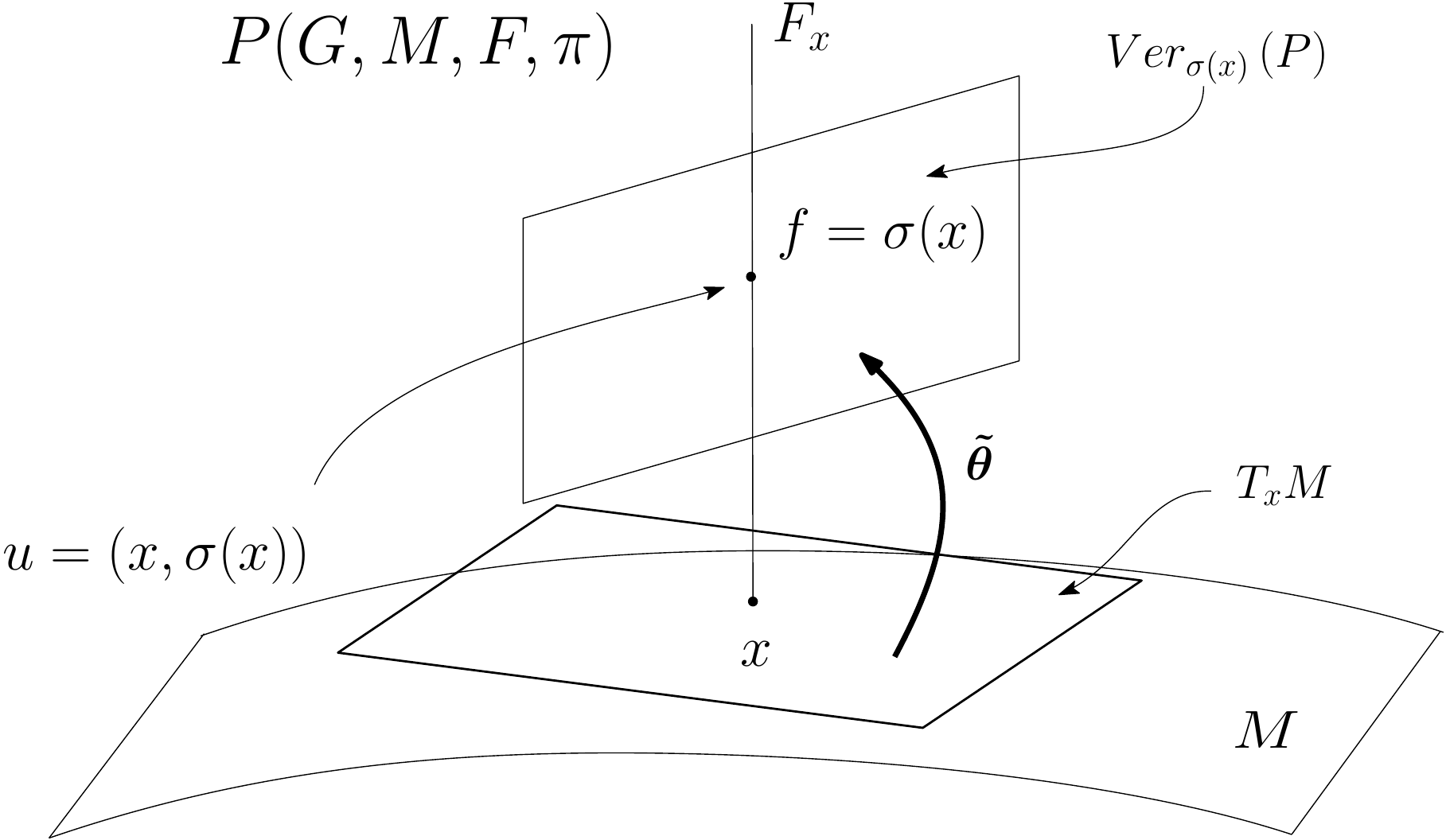}
 \caption{
 The solder form renders each tangent of the bundle base isomorphic to the corresponding vertical tangent space along the (global) section $\sigma$.
In a rigorous way, $\widetilde{\theta}$, the solder form, is the bundle map between $TM$ and $\sigma^*$Ver$(P)$.}\label{FIG-SolderFormOrigin}
 \end{center}
\end{figure}

Let  $P(G, M,  F, \pi)$ be a fiber bundle where $M$ is the base manifold, $G$ the (Lie) structure group, $F$ the fiber, 
$P$ the total space, and  $\pi$  the projection from $P$ onto $M$.
The definition of the solder form can be found in the original work of Kobayashi \cite{Kobayashi:1957}. Using the definition of the 
vertical bundle -- that is, the subbundle Ver$(P)$ of $TP$ defined as the disjoint union of the vertical spaces Ver$_pP$ for each $p$ in $P$ -- 
the definition given in \cite{Kobayashi:1957} reads:
\begin{quotation}
The bundle $P(G, M,  F, \pi)$ is soldered to $M$, if the following conditions are satisfied.
\begin{enumerate}
 \item $G$ is transitive on $F$.
 \item dim$F$ = dim$M$.
 \item $P(G, M,  F, \pi)$ admits a section $\sigma$ which will be identified with $M$.
\item There exists a
linear isomorphism of vector bundles $ \tilde \theta~:~TM \longrightarrow \sigma^*\mathrm{Ver}(P)$ from the tangent bundle of $M$ to the pullback of the vertical bundle of $P$  along the section $\sigma$.
\end{enumerate}
\end{quotation}
This last  condition can be interpreted as saying that $\tilde \theta$ determines a linear isomorphism
$$\tilde \theta_x : T_x M \longrightarrow V_{\sigma(x)} P $$
from the tangent space of $M$ at $x$ to the (vertical) tangent space of the fiber at the point $\sigma(x)$.  
This general definition is pictured in Fig.~\ref{FIG-SolderFormOrigin}.

The above general definition of soldering does not apply as is to principal bundles.  
Its application to a principal fiber bundle requires the use of
an intermediate associated vector bundle discussed in appendix \ref{App-AssocBundle}, as shown in Fig.\ref{FIG-SolderForm} since a global section in a principal bundle would otherwise render it trivial.
\begin{figure*}[th!]
\begin{center}
\includegraphics[width = 1\textwidth]{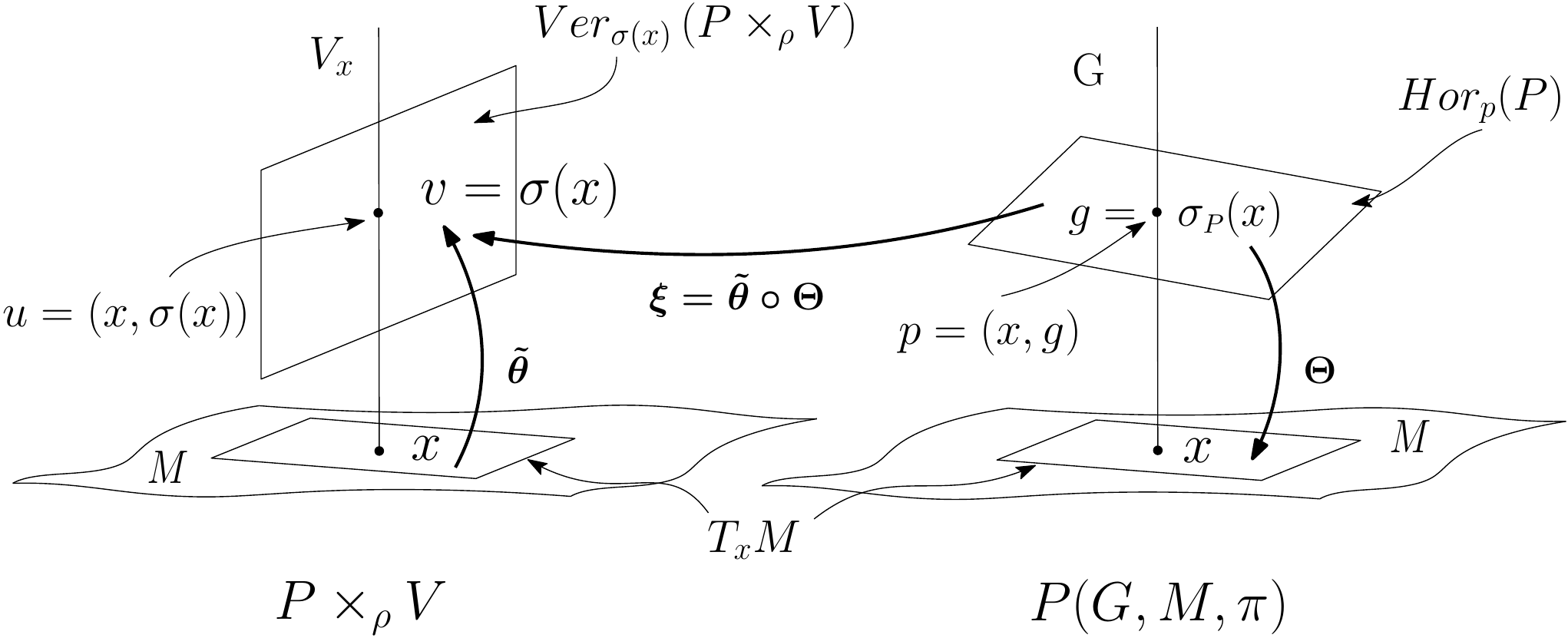}
 \caption{The solder form renders a tangent of the principal bundle base isomorphic to the vertical of an associated vector bundle. For a principal bundle, soldering is realised through isomorphism $\xi$ between one principal bundle's horizontal and the associated bundle's vertical at the corresponding section. This is made possible through the isomorphism $\Theta$ that always exists between the tangent to the base and any horizontal.}\label{FIG-SolderForm}
 \end{center}
\end{figure*}%
There, the soldering of the principal bundle is realised through the isomorphism $\xi$ between a horizontal and the associated vector bundle's vertical tangent, effected thanks to the isomorphism $\Theta$ between any horizontal of the principal bundle and the tangent to the base. In the case of the 
Frame bundle, as seen in Sec.~\ref{SEC-1-PreliminaryEheresmannSolderForm}, the function of the isomorphism $\Theta$ is played by the canonical form $\theta$. 

\begin{figure*}
\begin{center}
\includegraphics[width = .95\linewidth]{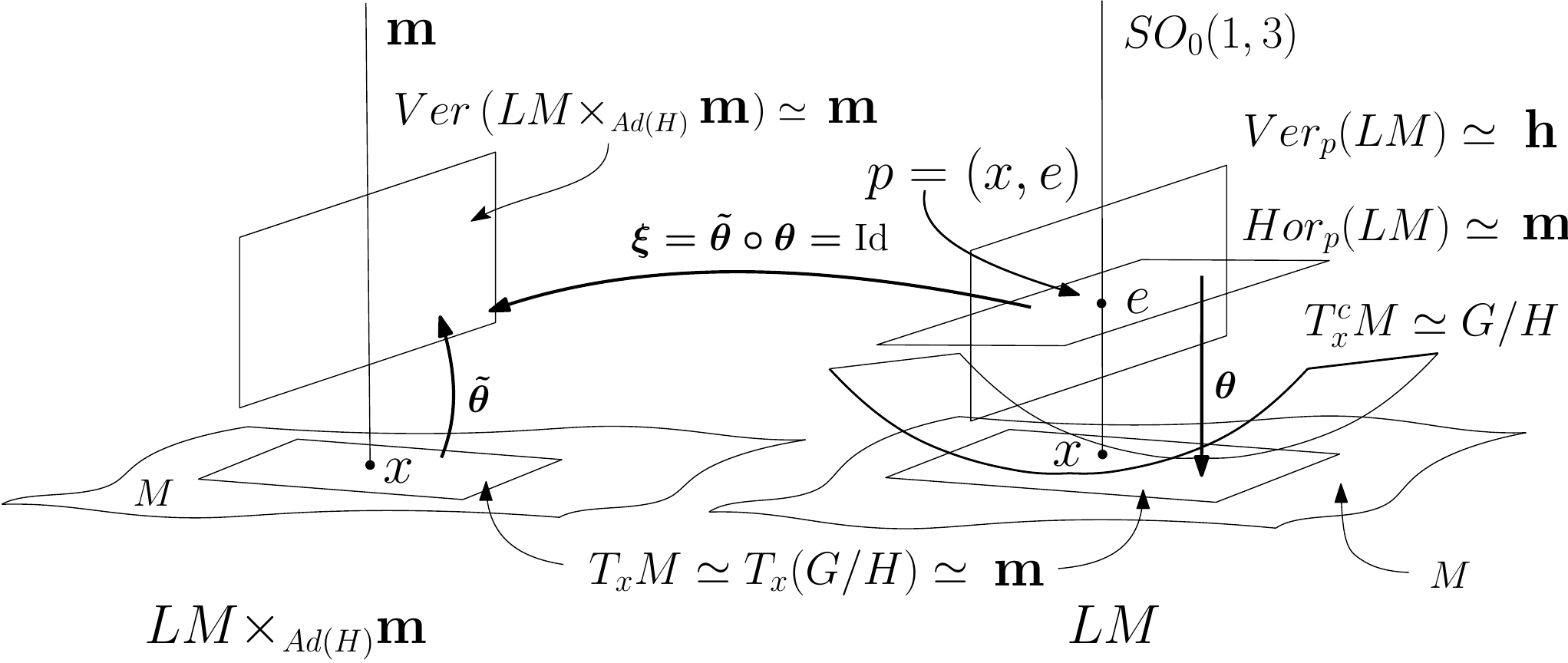}
\caption{The solder form $\widetilde{\theta}$ in the frame bundle $LM$ within Cartan geometry is the identity between the base's tangent planes and the vertical spaces of the associated vector bundle. Here the Cartan geometry is assumed to be reductive, so that one has the Ad$(H)$-invariant splitting
of the Lie algebra $\mathfrak{g} \equiv Lie(G) = \boldsymbol{h} \oplus \boldsymbol{m}$ with $\boldsymbol{h} = Lie(H)=\mathfrak{h}$ and 
$\boldsymbol{m} = \mathfrak{g/h}$.}\label{FIG-CartanReduc}
\end{center}
\end{figure*}

In the case of Cartan reductive geometry, soldering takes a specific form due to the various isomorphisms in which 
the $\mathfrak{g}/\mathfrak{h}$ part of the splitting, related to the translational part $\theta$ of the connection,   
$\mathfrak{g} \simeq \mathfrak{h} \oplus \mathfrak{g}/\mathfrak{h}$ of $\mathfrak{g}$ as Ad$(H)$-module,  is involved 
(see Fig. \ref{FIG-CartanReduc}). Indeed, the 
soldering can further be particularized, through the precising of these isomorphisms  
\cite[see, in particular,][]{Catren:2014vza}.

\subsection{Associated (vector) bundle.}\label{App-AssocBundle}
\begin{figure}[th!]
\begin{center}
\includegraphics[width = 8.5cm]{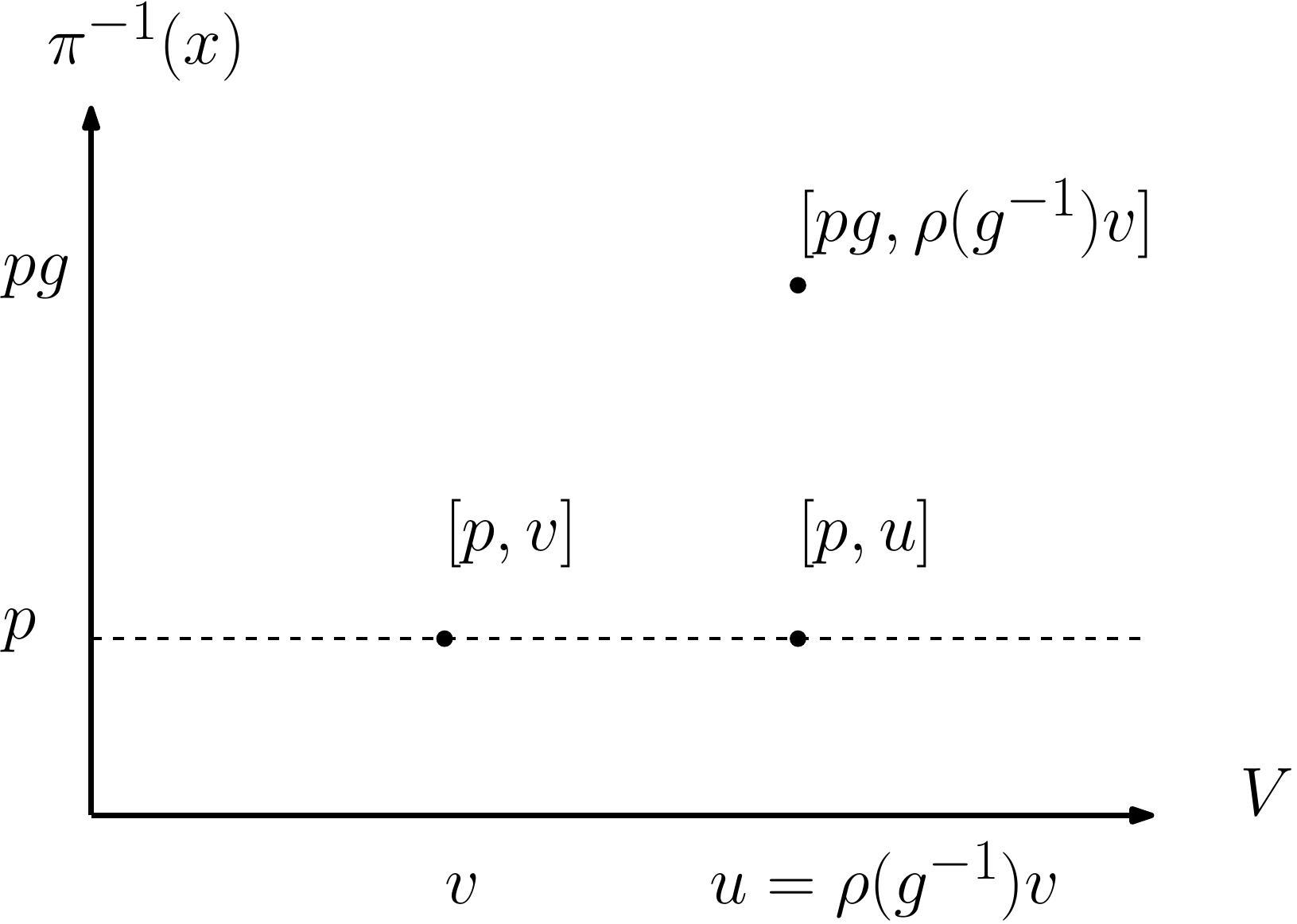}
 \caption{Linear structure in $P\times_\rho V$.}\label{FIG-TegrNoteIsoVf}
 \end{center}
\end{figure}

We first follow here \cite[p.54-55]{KobayashiNomizu:1963}  and then make some remarks. 

The idea is to build a fiber bundle in which the fiber $G$ of the original principal $G$-bundle is 
replaced by a manifold $F$  on which the group $G$ acts on the left. To this end, one defines a right action of the 
group $G$ on the product space $P\times F$ as follow: a  $g \in G$ maps $(p, f) \in P\times F$  to 
$(pg, g^{-1}f) \in P\times F$. The  set of the orbits (that is the equivalence classes) corresponding to this action 
is denoted $P\times_{\sss G} F$ and is the total space of the associated bundle. At first, $P\times_{\sss G} F$ 
is just a set, the structure of a fiber bundle is obtained as follow. One considers the mapping:
$(p ,f) \in P\times F \mapsto \pi(p) = x \in M$. It induces a projection $\hat \pi$ from $P\times_{\sss G} F$ 
onto the base $M$. The fiber of 
$P\times_{\sss G} F$ over $x\in M$ is $\hat\pi^{-1}(x)$. Now, in a neighborhood $U$ of $x$,  
$\pi^{-1}(U)\sim U\times G$ the action
of $G$ on $\pi^{-1}(U)\times F$ is: $(x, g', f) \mapsto (x, g'g, g^{-1}f)$ with 
$(x, g', f) \in U\times G \times F$ and $g\in G$. The isomorphism
$\pi^{-1}(U)\sim U\times G$ induces an isomorphism $\hat\pi^{-1}(U) \sim  U\times F$. Then, one can 
introduce a differentiable structure 
to ensure that $\hat \pi$ is a differentiable mapping from $P\times_{\sss G} F$ to the base $M$. 
This in turn ensures that $P\times_{\sss G}F$
is a fiber bundle with base $M$, fiber $F$, and structure group $G$.  

When $F$ is a $k$ dimensional vector space $V$ on which $G$ acts on the left through a representation $\rho$,
one obtains an associated vector bundle denoted by its total space $P\times_\rho V$, the right action on 
$P\times V$ is, in that case,
$(p, v) \mapsto (pg, \rho(g^{-1})v)$. We now restrict to this case.

The first remark is to recall 
that this rather complicated procedure is done to obtain a fiber bundle in which the fiber of the principal bundle
is replaced by a vector space $V$, all other data (in particular transition functions) remain unchanged. 
It should be noted that the
right action defined on $P\times V$ is different from the usual left action $\rho(g)v$. 
One can be puzzled by the fact that although the total space $P\times_\rho F$ is made of orbits, the fiber is 
the vector space $V$. 
Indeed, there is an identification there: following Fecko \cite[Sec. 2.4.1]{Fecko:2006}  there is a 
non-canonical isomorphism between $\hat \pi^{-1}(x)$ and $V$, given by $v \mapsto [p, v]$ for an arbitrary 
but fixed $p$. Since the right action 
on the product space $P\times V$ moves both $v$ and $p$ the fact that $p$ is held fixed allows us to
distinguish between two elements $u, v$ of $V$ belonging
to the same orbit through $G$, that is $u = \rho(g^{-1}) v$ (see Fig. \ref{FIG-TegrNoteIsoVf}). 
Accordingly, a linear structure on each fiber is provided through 
$\hat\pi^{-1}(x)$ by $[p, v] + \lambda [p, u]  := [p, v+\lambda u]$.
Different choices of $p$ do not change the result (they  
correspond to chooseing another line parallel to the $V$ axis on Fig. \ref{FIG-TegrNoteIsoVf}).

\subsection{On the affine connection}\label{App-AffineConnection}

In the search for a correct geometrical description of a translation gauge theory, the affine group presents 
a natural way to combine the translation of diffeomorphisms with the linear group of all possible 
frame transformations, which can be restricted to the Lorentz group. From \cite[p136]{KobayashiNomizu:1963}, 
the affine frame bundle $AM$ can be built with 
an affine connection $\tilde{\omega}$ and the homomorphism $$\begin{array}{rrl}
\gamma: & GL(n;\mathbb{R})\rightarrow & A(n;\mathbb{R})\\
 & {\textstyle {a}\mapsto} & \left(\begin{array}{cc}
a & 0_{n}\\
0 & 1
\end{array}\right)
\end{array}$$ 
from the frame bundle $LM$, and such that 
$$\gamma^*\tilde{\omega}=\omega+\phi$$ where $\omega$ is the connection on $LM$ and $\phi$ is an $\mathbb{R}^n$ 
valued connection 1-form.
Defining as usual the curvature of some connection $\omega$ with the Cartan structure equation: 
\begin{equation*}
\Omega_\omega := D_\omega \omega :=  d\omega + \omega \wedge \omega,
\end{equation*}
$D_{\omega}$ being the covariant derivative associated to
the connection $\omega$, one obtains the relation
\begin{align}
\gamma^*\tilde{\Omega}=\Omega+ D_{\omega}\phi.
\end{align}
Noticing the formal identity between the covariant derivative of $\phi$ 
and Eq.~(\ref{EQ-DefTorsion}), 
we obtain precisely the sum of curvature and torsion required to obtain the field strength for a TEGR translation-bundle
\begin{align}
\gamma^*\tilde{\Omega}=\Omega_{\sss L}+\Theta.
\end{align}
However, since it is not a curvature per se, the field strength interpretation of the affine curvature pullback is problematic.


\end{document}